\documentclass[12pt]{article}
\usepackage{amsmath,amsfonts,amsthm,amssymb,amscd,colordvi}
\usepackage{color}
\usepackage{showkeys}
\renewcommand{\Re}{\mathop{\rm Re}\nolimits}
\renewcommand{\Im}{\mathop{\rm Im}\nolimits}
\newcommand{\p}{\partial}

\newcommand{\vp}{\varphi}

\newcommand{\gi}{\rho}
\newcommand{\la}{\lambda}
\newcommand{\lla}{\gamma}

\newcommand{\bk}{{\mathbf k}}

\newcommand{\bb}{\mbox{\boldmath$\beta$}}
\newcommand{\R}{{\mathbb R}}

\newcommand{\Z}{{\mathbb Z}}

\newcommand{\E}{{\bf E}}

\newcommand{\T}{{\mathbb T}}
\newcommand{\N}{{\mathbb N}}

\newcommand{\cD}{{\cal D}}

\newcommand{\cH}{{\cal H}}

\newcommand{\cHR}{{\cal H}^{\text{res}}}

\newcommand{\cM}{{\cal M}}

\newcommand{\cT}{{\cal T}}

\newcommand{\strela}{\rightharpoonup}

\def\12{\tfrac12}

\def\eps{\varepsilon}
\theoremstyle{plain}
\newtheorem{theorem}{Theorem}[section]

\theoremstyle{definition}

\theoremstyle{remark}

\numberwithin{equation}{section}

\title{Derivation of the Kolmogorov-Zakharov equation  from
  the resonant-averaged stochastic NLS equation}

\author{Sergei
Kuksin\footnote{CNRS and  I.M.J, Universit\'e Paris Diderot-Paris 7, Paris, 
 France, e-mail:
  kuksin@math.jussieu.fr },\addtocounter{footnote}{2} Alberto Maiocchi \footnote{Laboratoire de
Math\'ematiques,  Universit\'e de Cergy-Pontoise, 2  avenue Adolphe Chauvin,
Cergy-Pontoise, France,
e-mail: alberto.maiocchi@unimi.it}}
\date{}

\begin{document}

\maketitle

\begin{abstract}
We suggest a new derivation of  a kinetic equation of
Kolmogorov-Zakharov (KZ) type for the spectrum 
of the weakly nonlinear Schr\"odinger equation with stochastic
forcing. The kynetic equation is obtained as a result of a double limiting
procedure. Firstly, we consider the equation on a finite box with periodic
boundary conditions and send the size of the nonlinearity and of the forcing
to zero, while the time is correspondingly rescaled; then, the size of
the box is sent to infinity (with a suitable rescaling of the
solution). We report here the results of the first limiting procedure,
analyzed with full rigour in \cite{KM13},  and show how the second
limit leads to a kinetic equation for the
spectrum, if some further hypotheses (commonly employed in the weak
turbulence theory) are accepted. Finally we show how to derive from
these equations the KZ spectra. 
\end{abstract}


\section{Introduction}
\subsection{Weak turbulence and spectra}
The theory of the weak turbulence studies weakly nonlinear PDEs, focusing
in particular on the distribution and exchange of energy among the
normal modes of oscillations (in most cases, these are  Fourier modes of the
solutions). The weakness of the nonlinearity allows to consider
the interaction between different modes (or, in a more physical
language, different waves) as a small perturbation to the linear flow,
so that the solutions of the equations can be approximated by suitable  power
series expansions. Usually, the lowest order nontrivial approximation
for the solution is considered and the attention is payed to its statistical properties
 on long time intervals. That is, one deals with averaged values of  certain 
quantities, taken with respect to some probability measure. The
latter can be introduced either as the probability of a given
configuration of initial data, or as the probability of a realization
of a stochastic forcing which is added to the system, together with a
damping to dissipates the energy pumped by the forcing.

The most important object of study is the distribution of energy among
the modes, that is the energy spectrum
$$
n_{\bk}(t)=\E\left[|v_{\bk}(t)|^2\right]\ ,
$$
where $v_{\bk}(t)$ denotes the amplitude of  ${\bk}$-th mode (${\bk}\in \R^d $ or ${\bk}\in\Z^d$) 
 at time $t$ and
$\E$ is the expected  value with respect to the probability
measure. Most of the predictions of the weak turbulence concern the
behaviour of $n_{\bk}$ as a function of ${\bk}$. Possibly the  most remarkable result
is the discovery of the existence of stationary solutions with spectra
decaying as a power law. In the case when the system is stirred  by
a forcing acting significantly only on some modes (think, for
instance, of the set $|{\bk}| \le r_1$) and is subject to a dissipation
having sensible effects only on a set of 
modes well separated from the first (for example, $|{\bk} |\ge r_2$, with
$r_2\gg r_1$), these solutions correspond to a constant flux of a
quantity (typically, the energy) among the modes $\bk$ such that   
$|\bk|\in [r_1,r_2]$, the so-called inertial zone. One then  says that a cascade
(of energy, etc.) occurs.

\subsection{Kolmogorov-Zakharov equation}
The main tool used to study the spectra is the kinetic equation (or,
better the class of equations) which
bears the names of Kolmogorov and Zakharov (KZ). It can be written as
$$
\frac{d}{dt} n_{\bk}= f_{\bk}(\{n_{\cdot}\})\ ,
$$
in which $f_{\bk}$ is a function of the whole spectrum $\{n_{\bk}, \bk\in\Z^d \}$, constructed 
in terms of the nonlinear  PDE. If one manages to derive the KZ
equation through some approximation procedures, then the problem of finding
stationary spectra reduces to that of finding spectra which make $f$
vanish (see \cite{EV} for the study of the Cauchy problem for the KZ
equation obtained from cubic nonlinear Schr\"odinger equation).

However, known derivations of the KZ equation all are heuristic, and 
serious doubts always existed concerning their validity. 
These concerns  have become even more serious
after the appearance of some results which seem to be in contradiction
with the prediction of the weak turbulence (see, for instance \cite{Majda,Mordant}).

In the  work \cite{KM13}  we propose a way to study the weak turbulence and
verify its postulates, in the frame of stochastic PDEs. The scheme
consists in considering a nonlinear PDE with stochastic forcing on a
torus of size $L$ and performing, in sequence,\footnote{The choice of
  the order of the limits is not that commonly considered in the
  theory of the weak turbulence.} two limiting procedures:
\begin{itemize}
\item the limit (on long times) in which the nonlinearity goes to zero
  together with the forcing; 
\item  the limit for $L\to \infty$ (with a possible scaling of the size of the solutions).
\end{itemize}

In \cite{KM13} the first limit is performed, using the method of resonant 
stochastic averaging in the spirit of Khasminskii, 
and it is proved that the
evolution of the spectrum on long times is governed by an
\emph{effective equation}, which contains only the resonant terms of
the nonlinearity. We intend here to show how, in the limit  $L\to
\infty$, such an equations leads one to obtain the KZ equation,
\emph{if some approximation is accepted, which are commonly used in
  the weak turbulence theory}.  In difference  with paper \cite{KM13}, we will
not care of mathematical rigour, focusing the interest on the possible
relevance of the results of \cite{KM13} for the deduction of KZ
equation, and thus for the weak turbulence.

\section{The limit of the weak nonlinearity and the effective equation}   
Now we   briefly sum up  the main results in
\cite{KM13}. There it was considered the 
Schr\"odinger equation on the torus $\T^d_L= \R^d/(2\pi L\Z^d)$,
\begin{equation}\label{*1}
u_t(t,x)-i\Delta u(t,x)=0,\quad x\in\T^d_L\,, 
\end{equation}
stirred by a perturbation, which comprises a
Hamiltonian  term, a linear damping and a random force. That is, we
have considered the equation 
\begin{equation}\label{1.11}
\begin{split}
u_t-i\Delta u= - i\eps^{2q_*}|u|^{2q_*}u-\nu f(-\Delta) u
+\sqrt\nu \frac{d}{dt}\sum_{\bk\in\Z^d_L}  b_\bk\bb^\bk(t) e^{i{\bk}\cdot x}
\ ,\\
 u=u(t,x),\quad x\in \T^d_L\,,
\end{split}
\end{equation}
where   $\ q_*\in \N $ and
$  \eps,\nu>0$ are two small parameters, controlling  the size of the
perturbation,  while $\Z^d_L$ denotes the set of vectors of the form
$\bk={\mathbf{l}}/L$ with ${\mathbf{l}}\in \Z^d$. The damping
$-f(-\Delta) $ is the selfadjoint  
  linear operator in $L_2(\T^d_L)$ which acts on the exponents $e^{i{\bk}\cdot x}$, $\bk\in\Z^d_L$, 
 according to
\begin{equation}\label{f}
f(-\Delta) e^{i{\bk}\cdot x}= \lla_\bk e^{i{\bk}\cdot x}, \qquad \lla_\bk= f( \lambda_\bk  )
\quad\text{where}\quad \lambda_\bk= |\bk|^2.
\end{equation}
 The function $f$ is real positive and continuous. To avoid technicalities,  we assume that 
$\ 
f(t)\ge C_1 |t| +C_2$ for all $t$,
for suitable positive constants $C_1,C_2$ (for example, $f(-\Delta)u=-\Delta u+u$). 
The processes $\bb^\bk, \bk\in\Z^d_L$, are
standard independent complex 
Wiener processes. The real numbers $b_\bk$ are all non-zero and decay  fast
when $|\bk|\to\infty$.

Equation \eqref{1.11} with small $\nu$ and $\eps$  is important for
physics and mathematical physics, where it serves  as a universal
model, see  \cite{Fal, Naz, ZL75, ZLF}. 
 The parameters $\nu$ and $\eps$ measure, respectively, the inverse 
 time-scale of the forced oscillations under consideration 
 and their  amplitude. We  consider the regime in which
  $$
  \eps^{2q_*}=\gi \nu,
  $$
  where $\rho>0$ is a constant. This assumption is in agreement with
  the requirement that one should consider the
  dynamics on a time-scale which becomes longer and longer as the
  amplitude goes to zero (see  \cite{ Naz} for the actual bounds
  imposed to the relation between $\eps$ and $\nu$). Passing to the slow
  time $\tau=\nu t$ and writing $u(\tau,x)$ as Fourier series,
$\ 
u(\tau,x)=\sum_\bk v_\bk(\tau)e^{i\bk\cdot x},
$ we  get the system
\begin{equation}\label{1.100}
\begin{split}
\dot v_\bk+i \nu^{-1}\lambda_\bk v_\bk=-\gamma_\bk v_\bk + 2\rho\, i   \,\frac{\p \cH(v)}{\p \bar v_\bk}  
+b_\bk \dot\beta^\bk(\tau),\quad \bk \in \Z^d_L.
\end{split}
\end{equation}
Here $v_{\bk}=v_{\bk}(\tau)$, the dot   $\ \dot{}\ $  stands for 
$\frac{d}{d\tau}$,  and   $\cH(v)$ is the Hamiltonian of the nonlinearity,
expressed in terms  of the Fourier coefficients 
$v=(v_\bk, \bk\in\Z^d_L)$:
\begin{equation}\label{Hv}
\cH(v)=\frac1{2q_*+2}
\sum_{\bk_1,\dots \bk_{2q_*+2}\in\Z^d_L} v_{\bk_1}\dots v_{\bk_{q_*+1}} \bar v_{\bk_{q_*+2}} \dots
\bar v_{\bk_{2q_*+2}} \,  \delta^{1\ldots q_*+1}_{q_*+2\ldots 2q_*+2}\,,
\end{equation}
where we used the notation (see \cite{Naz}):
\begin{equation}\label{N1}
\delta^{1\ldots q_*+1}_{q_*+2\ldots 2q_*+2}=\left\{\begin{array}{cc}
1 & \mbox{if }\bk_1+\ldots+\bk_{q_*+1}-\bk_{q_*+2}-\ldots-\bk_{2q_*+2} = 0 \\
0 & \mbox{otherwise}
\end{array}
\right.\ .
\end{equation}
As before  we are interested in  the limit $\nu\to0$, corresponding
to small oscillations in the original non-scaled equation.

The limiting procedure rests on the stochastic averaging theorem for
resonant systems with an infinite number of degrees of freedom (see
Introduction in \cite{KM13}). Let us consider the  equation
\begin{equation}\label{*eff1}
\begin{split}
\dot v_\bk=-\gamma_\bk v_\bk + 2\rho\, i   \,\frac{\p \cH^{\text{res}}(v)}{\p \bar v_\bk}  
+b_\bk \dot\beta^\bk(\tau),\quad \bk\in\Z^d_L,
\end{split}
\end{equation}
where $\cHR$ is obtained as the resonant average of the Hamiltonian $\cH(v)$: 
\begin{equation}\label{hres}
\cHR(v)= \frac1{2q_*+2}
\sum_{\bk_1,\dots \bk_{2q_*+2}\in\Z^d_L} v_{\bk_1}\dots v_{\bk_{q_*+1}} \bar v_{\bk_{q_*+2}} \dots
\bar v_{\bk_{2q_*+2}} \,  \delta^{1\ldots q_*+1}_{q_*+2\ldots 2q_*+2}\,
 \delta(\lambda^{1\ldots q_*+1}_{q_*+2\ldots 2q_*+2})\,,
\end{equation}
and we use another standard notation:
\begin{equation}\label{N2}
\delta(\lambda^{1\ldots q_*+1}_{q_*+2\ldots 2q_*+2})=\left\{\begin{array}{cc}
1 & \mbox{if }\lambda_{\bk_1} +\ldots+\lambda_{\bk_{q_*+1}}
-\lambda_{\bk_{q_*+2}}- \ldots- \lambda_{\bk_{2q_*+2}}
 = 0 \\
0 & \mbox{otherwise}
\end{array}
\right.\ .
\end{equation}
That is,  equation \eqref{*eff1} is obtained from the system 
 \eqref{1.100} by a simple procedure: we remove fast  terms $i\nu^{-1}\lambda_\bk v_\bk$ and
 replace the Hamiltonian $\cH$  by its resonant average $\cHR$.

If $v(\tau)$ is a solution of \eqref{1.100} or \eqref{*eff1} and $I(v(\tau))$ is its 
 vector of  energies,
$I(v(\tau))=\{|v_\bk(\tau)|^2,  \bk\in \Z^d_L\}$, then the main result
of \cite{KM13} is the following theorem, which shows that the
evolution of the energy vector is given by the effective equation
\eqref{*eff1}. 
 Moreover, in the stationary regime the effective equation completely
controls the limiting distribution of solutions:

\begin{theorem}\label{teor:1}
 Let $v_0:= \{v_{0_\bk}, \bk \in \Z^d_L\}$ be a
sufficiently smooth initial condition and $v^\nu(\tau)$ be a solution
for \eqref{1.100} with $v^\nu_0=v_0$. When $\nu\to0$, we have the
weak convergence of  measures  
$$
\cD(I(v^\nu(\tau)))\strela  \cD(I(v^0(\tau)))\,, \qquad 0\le\tau\le T\,,
$$
where $v^0(\tau)$  is a  solution of equation 
\eqref{*eff1} such that $v^0(0)=v_0$. 

Moreover, if equation \eqref{*eff1} has a unique stationary measure $\mu^0$ \footnote
{This happens e.g. if $q_*=1$, $d\le 3$, or id $d$ is any, but the function $f(\lambda)$
growth with $\lambda$ sufficiently fast,  see \cite{KM13}.}
and $v^\nu(\tau)$ is a stationary (in time)  solution of eq.~ \eqref{1.100},
 then
$$
\cD(v^\nu(\tau))\strela  \mu^0\ .
$$
\end{theorem}

Motivated by this result, we call \eqref{*eff1} the {\it effective equation} 
(for \eqref{1.100}).

\section{The limit  $L\to \infty$}
From the point of view of mathematics, the limit in which the size $L$
of the torus tends to infinity in equation \eqref{*eff1} (which, as we
have seen, determines the spectrum) presents serious problems, in
particular for what concerns the definition of the stochastic forcing.
We prefer  instead  to study for finite $L$ some averaged quantities,
calculated for solutions of the equation  (i.e., their moments), and  then to
pass to the limit as $L\to\infty$ only for them. 

For the sake of simplicity, we will consider the case of cubic
nonlinearity, i.e., we choose $q_*=1$ in \eqref{1.11}. Then the
effective equation takes the form 
\begin{equation}
d v_\bk(\tau)= \left(-\lla_\bk v_\bk -i \gi \sum_{\bk_1,\bk_2,\bk_3\in
\Z^d_L}
v_{\bk_1}v_{\bk_2} \bar v_{\bk_3}\delta^{\bk_1\bk_2}_{\bk_3 \bk}
\delta(\la^{\bk_1\bk_2}_{\bk_3\bk})\right)d\tau + b_\bk d\bb_\bk\ , \quad \bk\in
\Z^d_L\ .\label{efficace} 
\end{equation}
We define the moment $M^{\bk_1\ldots
  \bk_{n_1}}_{\bk_{n_1+1}\ldots \bk_{n_1+n_2}}$ of order $n_1+n_2$ as
\begin{equation}\label{moments}
M^{\bk_1\ldots \bk_{n_1}}_{\bk_{n_1+1}\ldots \bk_{n_1+n_2}}(\tau)=\E_\tau\left(
    v_{\bk_1} \cdots v_{\bk_{n_1}} \bar v_{\bk_{n_1+1}}\cdots
     \bar v_{\bk_{n_1+n_2}}\right) \ ,
\end{equation}
where $\E_\tau$ denotes the expected values at time $\tau$, i.e.,
$\E_\tau[f(v)]=\E[f(v(\tau))]$ for any function $f(v)$. 
Note  that $M_{\bk_1\ldots \bk_{n_1}}^{\bk_{n_1+1}\ldots \bk_{n_1+n_2}}= \bar
M^{\bk_1\ldots \bk_{n_1}}_{\bk_{n_1+1}\ldots \bk_{n_1+n_2}}$.
In order to write the evolution equation for the moments we put
$$
A M^{\bk_1\ldots \bk_{n_1}}_{\bk_{n_1+1}\ldots \bk_{n_1+n_2}} =-\left(\sum_{l=1
}^{n_1+ n_2}  \lla_{\bk_l}\right)  M^{\bk_1\ldots   \bk_{n_1}}_{\bk_{n_1+1}\ldots \bk_{n_1+n_2}}  \ , 
$$
and introduce the operator $\Gamma_{\bk_l}$ which erases the
index $\bk_l$ (in lower or upper position) according to
$$
\Gamma_{\bk_l} M^{\bk_1\ldots   \bk_{n_1}}_{\bk_{n_1+1}\ldots \bk_{n_1+n_2}}  =
\left\{ \begin{array}{cc}
M^{\bk_1\ldots\not \bk_l\ldots   \bk_{n_1}}_{\bk_{n_1+1}\ldots \bk_{n_1+n_2}} &
{if }\quad  l\le n_1\\
M^{\bk_1\ldots  \bk_{n_1}}_{\bk_{n_1+1}\ldots\not \bk_l\ldots  \bk_{n_1+n_2}} &
{if } \quad l> n_1 
\end{array}\right. \ .
$$
So, by making again  use of Ito's  formula, we get
\begin{equation}\label{eq:chain}
\begin{split}
\frac{d M^{\bk_1\ldots \bk_{n_1}}_{\bk_{n_1+1}\ldots \bk_{n_1+n_2}}}{d \tau} =&
A M^{\bk_1\ldots \bk_{n_1}}_{\bk_{n_1+1}\ldots \bk_{n_1+n_2}}\\
&-i\gi \Biggl( 
\sum_{l=1}^{n_1} \sum_{\bk'_1,\bk'_2,\bk'_3}
\Gamma_{\bk_l} M^{\bk_1\ldots \bk_{n_1}\bk'_1\bk'_2}_{\bk_{n_1+1}\ldots
  \bk_{n_1+n_2}\bk'_3} \delta^{\bk'_1\bk'_2}_{\bk'_3\bk_l} \delta(\la^{\bk'_1\bk'_2}_{\bk'_3\bk_l}) 
\\
&-
\sum_{l=n_1+1}^{n_1+n_2} \sum_{\bk'_1, \bk'_2\bk'_3}
\Gamma_{\bk_l}  M^{\bk_1\ldots \bk_{n_1}\bk'_3}_{\bk_{n_1+1}\ldots
  \bk_{n_1+n_2}\bk'_1\bk'_2} \delta_{\bk'_1\bk'_2}^{\bk'_3\bk_l} \delta(\la_{\bk'_1\bk'_2}^{\bk'_3\bk_l})
\Biggr)\\
&+2\sum_{l=1}^{n_1}\sum_{m=n_1+1}^{n_1+n_2}b^2_{\bk_l} \delta^{\bk_l}_{ \bk_{m}}
\Gamma_{\bk_l}\Gamma_{\bk_m}  M^{\bk_1\ldots \bk_{n_1}}_{\bk_{n_1+1}\ldots \bk_{n_1+n_2}}\ ,
\end{split}
\end{equation}
This equation expresses the moment of order $n_1+n_2$ as a function of
the moments of order $n_1+n_2-2$ and those  of order $n_1+n_2+2$. The
coupled system containing equations for all moments is called the
chain of moments equation (see \cite{MY}).\footnote{Notice that, due
  to our choice of 
  the degree of the nonlinearity, in our case the equations for
  moments of even order are decoupled from those for moments of odd
  order.} Systems of this kind are usually treated  by
approximating  moments of high order by suitable functions of lower order moments in
order to get a closed system of equations. We will show that if the 
quasi-Gaussian approximation (see below) 
 is chosen to close the system of moment equations,
then under the limit $L\to \infty$  we recover  a modified version of the KZ equation. 

We start from \eqref{eq:chain} for $M_\bk^\bk$ and a fixed $L$, which gives
\begin{equation}\label{eq:chain_2}
\dot M^\bk_\bk=-2\lla_\bk M_\bk^\bk + 2 b^2_\bk +2\gi \sum_{\bk_1,\bk_2,\bk_3} \Im M^{\bk_1 \bk_2}_{\bk \bk_3}
\delta^{\bk_1\bk_2}_{ \bk \bk_3} \delta(\la^{\bk_1\bk_2}_{ \bk \bk_3})\ .
\end{equation}
To study the sum in the r.h.s., we notice if  the Kr\"onecker deltas
are different from zero  because $\bk$  equals to one among
$\bk_1,\bk_2$ and $\bk_3$ is equal to 
another, then the moment is real and does not
contribute to the sum. So we may assume that 
 $\bk\neq \bk_1,\bk_2$, $\bk_3\neq
\bk_1,\bk_2$. In this case to calculate  the fourth order moments 
 in the r.h.s. of \eqref{eq:chain_2} we consider corresponding equation
 \eqref{eq:chain} of order four. Due to the just mentioned restriction on $\bk, \bk_1-\bk_3$,
 the last double sum in the r.h.s. of  \eqref{eq:chain} vanishes, and we get:
\begin{equation}\label{eq:chain_4}
\begin{split}
\dot M^{\bk_1 \bk_2}_{\bk \bk_3}=& -(\lla_\bk+\lla_{\bk_1}+\lla_{\bk_2}+\lla_{\bk_3})
M^{\bk_1 \bk_2}_{\bk \bk_3} 
+i\gi\sum_{\bk_4,\bk_5,\bk_6} \Bigl(M^{\bk_1 \bk_2 \bk_4}_{\bk_3 \bk_5 \bk_6}
\delta^{\bk \bk_4}_{ \bk_5 \bk_6} \delta(\la^{\bk\bk_4}_{ \bk_5\bk_6})+ \\
&M^{\bk_1 \bk_2 \bk_4}_{\bk \bk_5 \bk_6}
\delta^{\bk_3 \bk_4}_{ \bk_5 \bk_6} \delta(\la^{\bk_3\bk_4}_{ \bk_5\bk_6})
 - M^{\bk_2 \bk_5 \bk_6}_{\bk \bk_3 \bk_4}
\delta_{\bk_1 \bk_4}^{ \bk_5 \bk_6} \delta(\la_{\bk_1\bk_4}^{ \bk_5\bk_6}) -  M^{\bk_1 \bk_5
\bk_6}_{\bk \bk_3 \bk_4}
\delta_{\bk_2 \bk_4}^{ \bk_5 \bk_6} \delta(\la_{\bk_2\bk_4}^{ \bk_5\bk_6} )\Bigr)  \ .
\end{split}
\end{equation}

We make now a first approximation by neglecting the term containing
the time derivative at the l.h.s. of \eqref{eq:chain_4}. This can be
justified, if $\tau$ is large enough, by the quasistationary
approximation (see also Section~2.1.3 in \cite{ZLF}).  Namely, let us write equation 
\eqref{eq:chain_4}  as
$$
\left(\frac{d}{d\tau}+(\lla_\bk+\lla_{\bk_1}+\lla_{\bk_2}+\lla_{\bk_3})
 \right) M^{\bk_1 \bk_2}_{\bk \bk_3}=f(\bk)\ .
$$
Notice that since all $\lla_\bk$'s are positive, then the linear differential equation 
in the l.h.s. is exponentially stable. Assume that $f(\bk)$ as a function of $\tau$ is
almost constant during time-intervals, sufficient for the relaxation of the differential 
equation. Then
$$
M^{\bk_1 \bk_2}_{\bk
  \bk_3}\approx\frac{f(\bk)}{\lla_\bk+\lla_{\bk_1}+\lla_{\bk_2}+
  \lla_{\bk_3}} \ .
$$
We can finally insert this in
\eqref{eq:chain_2}, we get 
\begin{equation}\label{eq}
\begin{split}
\dot M^\bk_\bk\approx &-2\lla_\bk M_\bk^\bk + 2 b^2_\bk +2\gi^2 \sum_{\bk_1,\bk_2,\bk_3}
\frac1{\lla_\bk+\lla_{\bk_1}+\lla_{\bk_2}+\lla_{\bk_3}} 
\delta^{\bk_1\bk_2}_{ \bk \bk_3} \delta(\la^{\bk_1\bk_2}_{ \bk \bk_3})\\
&\Re\left(\sum_{\bk_4,\bk_5,\bk_6} \Bigl(M^{\bk_1 \bk_2 \bk_4}_{\bk_3 \bk_5 \bk_6} 
\delta^{\bk \bk_4}_{ \bk_5 \bk_6} \delta(\la^{\bk\bk_4}_{ \bk_5\bk_6})+
M^{\bk_1 \bk_2 \bk_4}_{\bk \bk_5 \bk_6}
\delta^{\bk_3 \bk_4}_{ \bk_5 \bk_6} \delta(\la^{\bk_3\bk_4}_{ \bk_5\bk_6})\right.
 \\&\left.\phantom{\sum_{\bk_4,\bk_5,\bk_6}}- M^{\bk_2 \bk_5 \bk_6}_{\bk \bk_3 \bk_4}
\delta_{\bk_1 \bk_4}^{ \bk_5 \bk_6} \delta(\la_{\bk_1\bk_4}^{ \bk_5\bk_6}) -
 M^{\bk_1 \bk_5
\bk_6}_{\bk \bk_3 \bk_4}
\delta_{\bk_2 \bk_4}^{ \bk_5 \bk_6} \delta(\la_{\bk_2\bk_4}^{ \bk_5\bk_6} )\Bigr)\right)
\ .
\end{split}
\end{equation}

We then apply a second approximation, generally accepted in the weak turbulence 
(see \cite{ZLF, Fal, Naz})
 which enables us to transform
the previous equation to a closed equation for the second order
moments. This consists in the quasi--Gaussian approximation, i.e., the
assumption that the higher-order moments \eqref{moments} 
can be approximated by polynomials of the second-order moments, 
as if the random variables $v_\bk$ were independent complex Gaussian
variables. So, in particular,
\begin{equation}\label{eq:qgauss}
M^{{\mathbf{l}}_1{\mathbf{l}}_2{\mathbf{l}}_3}_{{\mathbf{l}}_4 {\mathbf{l}}_5{\mathbf{l}}_6}\approx
M_{{\mathbf{l}}_1}^{{\mathbf{l}}_1}
M_{{\mathbf{l}}_2}^{{\mathbf{l}}_2}
M_{{\mathbf{l}}_3}^{{\mathbf{l}}_3}(\delta_{{\mathbf{l}}_1}^{{\mathbf{l}}_4}+  
\delta_{{\mathbf{l}}_1}^{{\mathbf{l}}_5}+
\delta_{{\mathbf{l}}_1}^{{\mathbf{l}}_6})
(\delta_{{\mathbf{l}}_2}^{{\mathbf{l}}_4}+ 
\delta_{{\mathbf{l}}_2}^{{\mathbf{l}}_5}+
\delta_{{\mathbf{l}}_2}^{{\mathbf{l}}_6})
(\delta_{{\mathbf{l}}_3}^{{\mathbf{l}}_4}+ 
\delta_{{\mathbf{l}}_3}^{{\mathbf{l}}_5}+ \delta_{{\mathbf{l}}_3}^{{\mathbf{l}}_6})\ ,
\end{equation}
where we have neglected corrections which appear when $\mathbf l_1$ is equal
to $\mathbf l_2$ or $\mathbf l_3$, as they give a vanishing contribution for $L\to \infty$.

At this point we pass in equation \eqref{eq}, closed using the relation \eqref{eq:qgauss},
to the limit $L\to\infty$.  To do  it we have to pass to a limit in the sum $S$ in the r.h.s.
of \eqref{eq}, by replacing the summation by integration. It is not
hard to see that the sum in \eqref{eq} splits into a finite number of
sums like
$$
S_{\bk}=\sum_{\bk_1,\bk_2,\bk_3} F_{\bk}(\bk_1,\mathbf
k_2,\bk_3) \delta^{\bk_1\bk_2}_{\bk \bk_3}
\delta(\lambda^{\bk_1 \bk_2}_{\bk \bk_3})\ ,
$$
where $\vec k:= (\bk_1,\bk_2,\bk_3)\in \Z^{3d}=:\cM$. Denote
$$
\Sigma_{\bk}=\left\{\vec x= (\mathbf x_1,\mathbf x_2,\mathbf x_3)\in
\R^{3d}: \mathbf x_1+\mathbf x_2=\bk+\mathbf x_3,
|\mathbf x_1|^2+|\mathbf x_2 |^2=|\bk|^2+|\mathbf x_3|^2\right\}\ .
$$
This is a manifold of dimension $3d-d-1=2d-1$, smooth outside $0$,
which lies  outside $\Sigma_{\bk}$ if $\bk\neq 0$, and is a singular point
of $\Sigma_{\bk}$ if $\bk=0$. For any non-zero $\vec x\in
\Sigma_{\bk}$ denote by $\pi_{\bk}(\vec x)$ the tangent
space $T_{\vec x} \Sigma_{\bk}$,
and denote by $\rho^{2d-1} \vp_{\bk}(\vec x)$ the $(2d-1)$-area
of the intersection of $\pi(\vec x)$ with the $\rho$-cube, centered at
$\vec x$ (with the sides parallel to the axes of $\R^{3d}$). Clearly,
$\vp_{\bk}(\vec x)$ is a smooth function on $\Sigma_{\bk}$ outside zero, such that
\begin{equation}\label{area}
V_1\le \vp_{\bk}(\vec x)\le V_1(3d)^{d-1/2}\ ,
\end{equation}
where $V_1$ is the volume of the $1$-ball in $\R^{2d -1}$. Let us put
$\rho = L^{-1}$ and write $S$ as
$$
S_{\bk}=L^{2d-1} \sum_{\substack{\vec x\in \Sigma_{\mathbf
      k}\cap \cM\\ \vec x\neq 0}} \left( \frac{F_{\bk}(\vec
  x)}{\vp_{\bk}(\vec x)}\right) \vp_{\bk}(\vec x) \rho^{2d-1}\ .
$$
The r.h.s. is the Riemann sum for the integral
$\int_{\Sigma_{\bk}\backslash\{0\} } \frac{F_{\bk}(\vec
  x)}{\vp_{\bk}(\vec x)} d\vec
x$. So 
$$
S_{\bk}\approx L^{2d-1} \int_{\Sigma\backslash \{0\}}
\frac{F_{\bk}(\vec
  x)}{\vp_{\bk}(\vec x)} d \vec x\ ,
$$
where $\vp_{\bk}(\vec x)$ is a smooth function satisfying \eqref{area}.

 After some calculations we get the limiting (as $L\to\infty$) equation in the form
\begin{equation*}
\begin{split}
\dot M^\bk_\bk\approx &-2\lla_\bk M_\bk^\bk + 2 b^2_\bk +2\gi^2L^{2d-1}
\int_{\R^{3d}} d\bk_1 d \bk_2 d
\bk_3  \frac{\vp_\bk^{-1}(\bk_1,\bk_2,\bk_3)}{\lla_\bk+\lla_{\bk_1}+\lla_{\bk_2}+\lla_{\bk_3}}
\delta^{\bk_1\bk_2}_{ \bk \bk_3} \delta(\la^{\bk_1\bk_2}_{ \bk \bk_3})\\
&\Bigl( M_{\bk_1}^{\bk_1} M_{\bk_2}^{\bk_2} M_{\bk_3}^{\bk_3}+ M_{\bk}^{\bk}
M_{\bk_1}^{\bk_1} M_{\bk_2}^{\bk_2}
- M_{\bk}^{\bk} M_{\bk_2}^{\bk_2}
M_{\bk_3}^{\bk_3}- M_{\bk}^{\bk} M_{\bk_1}^{\bk_1} M_{\bk_3}^{\bk_3}\Bigr)\ .
\end{split}
\end{equation*}

Finally, we consider the quantities  $n_\bk=L^d M_\bk^\bk/2$ (so that $\sum_\bk
M_\bk^\bk/2= \int n_\bk$ in the  limit when $L$ going to infinity), 
 define $\tilde b_\bk=L^{d/2} b_\bk$ and put $\gi=\epsilon^2L^{1/2}
$ , getting 
\begin{equation}\label{eq:kz_1}
\begin{split}
\dot n_\bk= -2\lla_\bk n_\bk+ \tilde b^2_\bk &+ 8\epsilon^4
\int_{\R^{3d}} d \bk_1\,d \bk_2\, d\bk_3
\delta^{\bk_1\bk_2}_{ \bk \bk_3} \delta(\la^{\bk_1\bk_2}_{ \bk \bk_3})
\frac{\vp_\bk^{-1}(\bk_1,\bk_2,\bk_3)}{\lla_\bk+\lla_{\bk_1}+\lla_{\bk_2}+\lla_{
    \bk_3}} 
\\
&\times \Bigl(n_{\bk_1}n_{\bk_2}n_{\bk_3}+ n_{\bk}
n_{\bk_1}n_{\bk_2}
- n_{\bk} n_{\bk_2}
n_{\bk_3}- n_{\bk} n_{\bk_1} n_{\bk_3}\Bigr)\ ,
\end{split}
\end{equation}
where $\vp_{\bk}(\vec x)$ is a smooth function satisfying \eqref{area}.
We have thus shown, that, with a proper scaling of $\rho$ and $b$, we
can get an equation which is very similar to the KZ equation for the
NLS (see, for instanc, formula (6.81) of \cite{Naz} for the $2-d$
case). The differences 
 are two: obviously in our case there appear the forcing and the
 dissipation, which were absent in his case; more interestingly, the denominator
$\lla_\bk+\lla_{\bk_1}+\lla_{\bk_2}+\lla_{\bk_3}$ appears in the integral,
which will modify the spectra.

\section{Kolmogorov-Zakharov spectra}
We show here how to deduce a power law spectrum from equation
\eqref{eq:kz_1}, following the well known Zakharov argument (see  \cite{ZLF,Naz}). 

First of all, we have to restrain our analysis  to the inertial
zone, i.e., to the spectral 
zone, where the  damping and the  forcing are negligible.
This means that we have to suppose that damping and forcing are such
that for wavevectors $\bk$ belonging to a sufficiently large spectral region
the first two terms at the r.h.s. of \eqref{eq:kz_1} can be
neglected. Thus, in the inertial zone we end up with the equation 
\begin{equation}\label{eq:kz_2}
\begin{split}
\dot n_\bk\approx &\,8 \epsilon^4
\int_{\R^{3d}} d \bk_1\,d \bk_2\, d\bk_3
\delta^{\bk_1\bk_2}_{ \bk \bk_3} \delta(\la^{\bk_1\bk_2}_{ \bk \bk_3})
\frac{\vp_\bk^{-1}(\bk_1,\bk_2,\bk_3)}{\lla_\bk+\lla_{\bk_1}+\lla_{\bk_2}+\lla_{
    \bk_3}} 
\\
&\times \Bigl(n_{\bk_1}n_{\bk_2}n_{\bk_3}+ n_{\bk}
n_{\bk_1}n_{\bk_2}
- n_{\bk} n_{\bk_2}
n_{\bk_3}- n_{\bk} n_{\bk_1} n_{\bk_3}\Bigr)\ ,
\end{split}
\end{equation}
Notice that, while we can simply approximate $\tilde b_\bk$ with zero,
this  cannot be done with $\lla_\bk$, as it appears in the
denominator of the integral at the r.h.s. of \eqref{eq:kz_1} (the
so-called collision term), and can play an essential role a
determination of the spectrum.

The previous equation has  the form of the four-wave kinetic
equation (see, for instance,  formula 2.1.29 of \cite{ZLF}). It is
well known (see \cite{ZLF,Naz} how to solve such an equation for
stationary spectra with the aid of Zakharov transformations, if the
terms 
$$
\cT^{\bk,\bk_3}_{\bk_1,\bk_2} =
\frac{\vp_\bk^{-1}(\bk_1,\bk_2,\bk_3)}{\lla_\bk+\lla_{\bk_1}+\lla_{\bk_2}+\lla_{
    \bk_3}} 
$$
satisfy some conditions of symmetry and homogeneity. Namely, one should
have that 
\begin{eqnarray*}
&\cT^{\bk,\bk_3}_{\bk_1,\bk_2}= \cT^{\bk_3,\bk}_{\bk_1,\bk_2}=
\cT^{\bk,\bk_3}_{\bk_2,\bk_1}=\cT^{\bk_1,\bk_2}_{\bk,\bk_3}\\
&\cT^{\lambda \bk,\lambda\bk_3}_{\lambda\bk_1,\lambda\bk_2}= \lambda^m
\cT^{\bk,\bk_3}_{\bk_1,\bk_2} \ ,
\end{eqnarray*}
for some $m\in \R$. For the
sake of simplicity, we confine ourselves to the isotropic  case when 
 $n_\bk$ is a function of $k=|\bk|$ only.

The requirements above are met  if $\vp$ can be approximated by a
constant and $\lla_\bk$ by a  homogeneous function of the form
$\lla_\bk=\eps |\bk|^m$, where $\eps_1\ll 1$ is a parameter that
guarantees that the dissipation term is indeed negligible and $m\in
\R$.\footnote{This agrees with the hypotheses on the dissipation (see
  \eqref{f}), if, for instance, we choose
$$
\lla_\bk= \eps_1+\eps_2 |\bk|^{\beta}\ ,
$$
where $\eps_1,\eps_2\ll 1$, and either $\eps_1\gg \eps_2$, which gives
$m=0$, or viceversa, which gives $m=\beta$.}

We operate (following \cite{Naz}, Sec. 9.2.2, see also \cite{ZLF},
Sec. 3.1.3) by integrating equation \eqref{eq:kz_2} over the
angles  and obtain that
$$
\dot n_k=8\epsilon^4 \int_0^{+\infty} \int_0^{+\infty} \int_0^{+\infty}
\mathfrak T^{k3}_{12}\, n_1 n_2 n_3 n_k (k_1 k_2 k_3)^{d-1} d k_1\, d
k_2\, d k_3\ ,
$$
where
$$
\mathfrak T^{k3}_{12}= \left(\frac1{n_k}+ \frac1{n_3}- \frac1{n_1} -
\frac1{n_2} \right)\delta(\la^{\bk_1\bk_2}_{
  \bk \bk_3}) \int \delta^{\bk_1\bk_2}_{ \bk \bk_3}
\cT^{\bk,\bk_3}_{\bk_1,\bk_2} d \Omega_1 d\Omega_2 d\Omega_3\ ,
$$
$C$ denotes a suitable positive constant and the integration is taken
on the $d$-dimensional solid angles $\Omega_i=\Omega(\bk_i)$. Due to
the symmetries of $\mathfrak T^{k3}_{12}$ (inherited  from those of
$\cT^{\bk,\bk_3}_{\bk_1,\bk_2}$),  by a proper renaming of mute integration variables
we can 
rewrite the previous equation as
\begin{equation}\label{eq:kz_3}
\dot n_k=\epsilon^4 \int
\left(\mathfrak T^{k3}_{12}+ \mathfrak T^{3k}_{12} - \mathfrak
T^{13}_{k2} - \mathfrak T^{23}_{1k}\right)  n_1 n_2 n_3 n_k (k_1 k_2
k_3)^{d-1} d k_1\, d 
k_2\, d k_3\\=: I_1+I_2+I_3+I_4. 
\end{equation}

We introduce then
as an ansatz the power law $n_k\propto k^\nu$, for real $\nu$, and, following Zakharov,
make in the integrals $I_2, I_3, I_4$ the following substitutions: in $I_2$ 
 we put
$$ 
k_1=\frac{k k'_1}{k'_3}\ ,\quad k_2=\frac{k k'_2}{k'_3}\ ,\quad
k_3=\frac{k^2}{k'_3} \ ,
$$ 
in $I_3$ put
$$
k_1=\frac{k^2}{k'_1}\ ,\quad k_2=\frac{k k'_2}{k'_1}\ ,\quad
k_3=\frac{k k'_3}{k'_1} \ ,
$$ 
and in $I_4$ --
$$
k_1=\frac{k k'_1}{k'_2}\ ,\quad k_2=\frac{k^2}{k'_2}\ ,\quad
k_3=\frac{k k'_3}{k'_2} \ .
$$ 
Then we re-denote the variables $k'_j$ back to $k_j$ and sum up the four 
integrals. By the homogeneity of $\cT^{\bk,\bk_3}_{\bk_1,\bk_2}$, one
gets that the integral in  
\eqref{eq:kz_3} is proportional to
\begin{equation}\label{collision}
\int \mathfrak T^{k3}_{12} \left[ 1+ \left(\frac{k_3}{k}\right)^x-
\left(\frac{k_1}{k}\right)^x - \left(\frac{k_2}{k}\right)^x\,\right] n_1
n_2 n_3 n_k (k_1 k_2 k_3)^{d-1} d k_1\, d 
k_2\, d k_3\ ,
\end{equation}
where
$$
x=2- 3\nu -m-3d\ .
$$

The stationary solutions are found by looking for $\nu$ for which the integral
in \eqref{collision} vanishes. In addition to the ``thermodynamical
equilibrium'' solutions $n_k=C$, $n_k=C/k^2$, two nontrivial
power law stationary distributions appear by equating to zero the term
in square brackets in \eqref{collision}, corresponding to $x=0$ and
$x=2$.\footnote{Notice that the second solution appears because the
  integration is restricted to the surface $k^2+k_3^2-k_1^2-k_2^2=0$,
  due to the term $\delta(\la^{\bk_1\bk_2}_{ \bk \bk_3})$.} These are
the Kolmogorov-Zakharov solutions: 
$$
n_k\propto k^{-(m+3d-2)/3}\ ,\qquad  n_k\propto k^{-(m+3d)/3}\ .
$$
They correspond to the well known Kolmogorov-Zakharov spectra for the
NLS equation  without dissipation (for $d\ge 2$) if $m=0$, while the
dissipation modify the power law of the decay if $m\neq 0$.

\bibliography{meas}
\bibliographystyle{amsalpha}

\end{document}